\documentclass[final]{svjour2}
\usepackage{graphicx}
\usepackage{rotating}
\usepackage{amssymb}
\usepackage{mathptmx}
\usepackage[numbers]{natbib}
\makeatletter
\journalname{Journal of Low Temperature Physics}

\bibpunct{}{}{,}{s}{}{,}

\begin{document}

\newcommand{\hdblarrow}{H\makebox[0.9ex][l]{$\downdownarrows$}-}
\title{Long-range Order in the A-like Phase of Superfluid 3He
in Aerogel}

\author{I.A.Fomin}

\institute{P. L. Kapitza Institute for Physical Problems, \\
ul. Kosygina 2, 119334 Moscow,Russia\\
Tel.:+7(495)1373248\\ Fax:+7(495)6512125\\
\email{fomin@kapitza.ras.ru}
\\2: XXXXXXXXX\\ 3: YYYYYYYYYYY}

\date{XX.XX.2007}

\maketitle

\keywords{Superfluid 3He, disorder and porous media}

\begin{abstract}
A mutual action of the random anisotropy brought in the superfluid
$^3$He by aerogel and of the global anisotropy caused by its
deformation is considered. Strong global anisotropy tends to
suppress fluctuations of orientation of the order parameter and
stabilizes ABM order parameter. In a limit of vanishing anisotropy
fluctuations of ABM order parameter became critical. It is argued
that still in a region of small fluctuations the order parameter
changes its form to be less sensitive to the random anisotropy.
For a favorable landscape of the free energy of superfluid $^3$He
the fluctuations remain small even in a limit of vanishing global
anisotropy and the long-range order is maintained.

PACS numbers: 67.57.-z, 67.57.Pq, 75.10.Nr
\end{abstract}

\section{Introduction}
Recent NMR experiments with the superfluid $^3$He in a uniaxially
compressed aerogel \cite{jap} have shown that the state of
 the A-like phase is very sensitive to a global anisotropy of
 aerogel induced by its deformation. The global anisotropy
 stabilizes long-range order in a contrast to the random local
 anisotropy which tends to disrupt this order.
The mechanism of disruption of a long-range order is the unlimited
growth of fluctuations of the order parameter in directions of its
degeneracy (Goldstone fluctuations) \cite{Lark,imry}. In the case
of superfluid $^3$He these are fluctuations of orientation of the
order parameter. Deformation of aerogel gives rise to the global
anisotropy which lifts degeneracy of the order parameter of
superfluid $^3$He with respect to the orbital rotations. The
lifting of degeneracy tends to suppress the Goldstone
fluctuations. So, the state of superfluid $^3$He in a deformed
aerogel is a result of competition between the random local and
regular global anisotropy. In a limit of strong global anisotropy
the Goldstone fluctuations are small and the order parameter of
the A-like phase has ABM form \cite{jap,dm1}. In a limit of
vanishing anisotropy, if the form of the order parameter is fixed
and only its orientation can vary, a possible result of disruption
of orientational long-range order is transition in the
Larkin-Imry-Ma (LIM) state \cite{vol_conf}. A straightforward
interpolation between the two limits does not exhaust
possibilities of variation of a state of $^3$He in aerogel with a
change of global anisotropy. There exist a feedback effect of
fluctuations on a form of the order parameter. Depending on a
landscape of the free energy of superfluid $^3$He in a vicinity of
the ABM order parameter this effect can be significant. Variation
of a form of the order parameter of $^3$He-A under the influence
of fluctuations adds another dimension to the manifold of possible
states of this phase. This possibility was considered previously
only for the isotropic aerogel \cite{fmn}. In the present paper
the global anisotropy is introduced in this analysis as an
external parameter.  It is shown in particular, that if both
anisotropy of aerogel and variation of a form of the order
parameter are taken into account, the long-range order in the
A-like phase of superfluid $^3$He can be preserved even in a limit
of a vanishing global anisotropy.

\section{Effect of anisotropy}
Interaction of aerogel with the superfluid $^3$He is described
phenomenologically by the extra term in the Ginzburg and Landau
functional:
$$
F_{\eta}=N(0)\int \eta_{jl}({\bf r})A_{\mu j}A_{\mu l}^*d^3r ,
\eqno(1)
$$
where $N(0)$ is the density of states at the Fermi level, $A_{\mu
j}$ -- the order parameter and $\eta_{jl}({\bf r})$ -- the random
anisotropy tensor. On the strength of $t\to -t$ invariance tensor
$\eta_{jl}({\bf r})$ is real and symmetric. For isotropic aerogel
the average $<\eta_{jl}({\bf r})>=0$. To account for a possible
global anisotropy of aerogel a constant (${\bf r}$-independent)
symmetric tensor $\kappa_{jl}$ has to be added to $\eta_{jl}({\bf
r})$.  The resulting expression for the GL free energy has the
following structure:
$$
 F_{GL}=N(0)\int d^3r[f_0+f_{\nabla}+
(\eta_{jl}({\bf r})+\kappa_{jl})A_{\mu j}A_{\mu l}^*]. \eqno(2)
$$
Here
$$
 f_0=\tau A_{\mu j}A_{\mu j}^*+
\frac{1}{2}\sum_{s=1}^5 \beta_sI_s       \eqno(3)
$$
is the unperturbed, or ``bare" GL free energy, $I_s$ - 4-th order
invariants in the expansion of the free energy over $A_{\mu j}$.
Coefficients $\beta_1,...\beta_5$ are phenomenological constants
\cite{VW}. Tensors $\eta_{jl}({\bf r})$ and $\kappa_{jl}$ can be
defined as traceless, i.e. their traces are included in the
definition of $\tau=(T-T_c)/T_c$.

For the gradient energy $f_{\nabla}$ we take a model isotropic
expression
$$
f_{\nabla}=\frac{2\xi_0^2}{5}\left(\frac{\partial A_{\mu
l}}{\partial x_n} \frac{\partial A^*_{\mu l}}{\partial
x_n}\right), \eqno(4)
$$
where $\xi_0=\hbar v_F/(2\pi T_c)$ is the coherence length in the
superfluid state. Equilibrium configuration of the order parameter
is found from the equation
$$
\frac{\partial f_0}{\partial A^*_{\mu
j}}-\frac{2\xi_0^2}{5}\left(\frac{\partial^2 A_{\mu j}}{\partial
x_n^2}\right)+\kappa_{lj}A_{\mu l} = -A_{\mu l}\eta_{lj}({\bf r})
\eqno(5)
$$
and its complex conjugated. For high porosity aerogel tensor
$\eta_{lj}({\bf r})$ can be treated as a small perturbation.
Solution of Eq.(5) can be sought as a sum of the average order
parameter $\bar A_{\mu j}$ and of a small fluctuation $a_{\mu
j}({\bf r})$:
$$
A_{\mu j}=\bar A_{\mu j}+a_{\mu j}({\bf r}).  \eqno(6)
$$
$\bar A_{\mu j}$ is assumed to be not far from one of the minima
of $f_0$. The long-range order exist when the average order
parameter is finite.

Following the standard perturbation procedure \cite{LarkOv} we
expand Eq. (5) up to the second order in $a_{\mu j}({\bf r})$ and
$\eta_{jl}({\bf r})$. The linear terms render equations for the
fluctuations:
$$
\frac{\partial^2f_0}{\partial A^*_{\mu j} \partial A_{\nu l}}
a_{\nu l}+\frac{\partial^2 f_0}{\partial A^*_{\mu j}\partial
A^*_{\nu l}} a^*_{\nu l}-\frac{2\xi_0^2}{5}\left(\frac{\partial^2
a_{\mu j}}{\partial x_n^2}\right)+\kappa_{lj}a_{\mu l} =
-\eta_{lj}\bar A_{\mu l}, \eqno(7)
$$
$$
\frac{\partial^2f_0}{\partial A_{\mu j} \partial A^*_{\nu l}}
a^*_{\nu l}+\frac{\partial^2 f_0}{\partial A_{\mu j}\partial
A_{\nu l}} a_{\nu l}-\frac{2\xi_0^2}{5}\left(\frac{\partial^2
a^*_{\mu j}}{\partial x_n^2}\right)+\kappa_{lj}a^*_{\mu l} =
-\eta_{lj}\bar A^*_{\mu l}, \eqno(8)
$$
and the average of Eq. (5) over the ensemble of $\eta_{jl}({\bf
r})$ -- the equation for the $\bar A_{\mu j}$:
$$
\frac{\partial f_0}{\partial A^*_{\mu
j}}+\frac{1}{2}\left[\frac{\partial^3f_0}{\partial A^*_{\mu j}
\partial A_{\nu l}\partial A_{\beta m}}<a_{\nu l}a_{\beta m}>+
2\frac{\partial^3f_0}{\partial A^*_{\mu j}
\partial A_{\nu l}\partial A^*_{\beta m}}<a_{\nu l}a^*_{\beta
m}>\right]+
$$
$$
<\eta_{jl}a_{\mu l}>+\kappa_{lj}\bar A_{\mu l}=0. \eqno(9)
$$
The average $<\eta_{jl}a_{\mu l}>$ can be combined with $\tau\bar
A_{\mu j}$ in $\frac{\partial f_0}{\partial A^*_{\mu j}}$. The
remaining averages of binary products of fluctuations i.e.
$<a_{\nu l}a_{\beta m}>=<a_{\nu l}({\bf r})a_{\beta m}({\bf r})>$
yield corrections to the order parameter.

 The state of the unperturbed superfluid $^3$He is continuously degenerate
with respect to separate rotations in spin and in orbital spaces.
The latter is of significance here. The random anisotropy
$\eta_{jl}({\bf r})$ breaks locally rotational degeneracy and
induces fluctuations $a_{\mu j}({\bf r})$. The ``longitudinal"
fluctuations, which change the magnitude and the form of the order
parameter are weakly effected by the global anisotropy. Their
binary averages were estimated before \cite{fom1}
$$
<a_{\mu j}a_{\nu n}>\sim\frac{1}{8\pi} \frac{\Phi_{jlmn}(0)}
{\xi_0^3}\frac{\bar A_{\mu l}\bar A_{\nu m}}{\sqrt{2|\tau|}}.
\eqno(10)
$$
Here
$$\Phi_{jlmn}(0)=
\left[\int<\eta_{jl}(\textbf{k})\eta_{mn}(-\textbf{k})>
\frac{do}{4\pi}\right]_{k=0}=\Phi_0(\delta_{jm}
\delta_{ln}+\delta_{jn}\delta_{lm}-\frac{2}{3}\delta_{jl}\delta_{mn}).
$$
Integral in the square brackets is taken over the solid angle $do$
in $\textbf{k}$-space. The relative value of these fluctuations
with respect to the square of the average order parameter is
characterized by the parameter
$g_{\tau}=\Phi_0/(\xi_0^3\sqrt{|\tau|})$. For aerogel with the
radius of strands $\rho$ and the average distance between them
$\xi_a$ $g_{\tau}\sim\rho^2/(\xi_0\xi_a\sqrt{|\tau|})$, which is
small if the temperature $T$ is not too close to $T_c$.

Effect of fluctuations of orientation of the order parameter, or
 transverse  fluctuations does depend on a global anisotropy. Let
us start with a ``strongly" compressed aerogel when definitely
$\bar A_{\mu j}=A^{ABM}_{\mu j}$:
$$
A^{ABM}_{\mu j}=\Delta\frac{1}{\sqrt{2}}\hat d_{\mu}(\hat
m_j+i\hat n_j). \eqno(11)
$$
Here $d_{\mu}$ is a unit vector in spin space, $\bf m$ and $\bf n$
- two mutually orthogonal unit vectors in orbital space. In a
uniaxially compressed aerogel vector $\bf l=\bf m\times\bf n$ is
oriented along the direction of compression, which will be taken
as $z$-axis. Then tensor $\kappa_{jl}$ is diagonal, with the
components $\kappa_{xx}=\kappa_{yy}=-\kappa, \kappa_{zz}=2\kappa$,
$\kappa>0$. To obtain equation for the transverse fluctuations we
have to multiply Eq.(7) by $\frac{\partial\bar A^*_{\mu
j}}{\partial\theta_q}=e^{jqn}\bar A^*_{\mu n}$, where $e^{jqn}$ is
antisymmetric tensor,  Eq.(8) by $\frac{\partial\bar A_{\mu
j}}{\partial\theta_q}=e^{jqn}\bar A_{\mu n}$ and to sum the
obtained equations. Vector $\theta_q$ specifies infinitesimal
rotation of the order parameter. The resulting equation is
$$
\frac{\partial\bar A^*_{\mu j}}{\partial\theta_q}\kappa_{jl}a_{\mu
l}+\frac{\partial\bar A_{\mu
j}}{\partial\theta_q}\kappa_{jl}a^*_{\mu
l}-\frac{2\xi^2_0}{5}\frac{\partial^2}{\partial
x_n^2}\left(\frac{\partial\bar A^*_{\mu
j}}{\partial\theta_q}a_{\mu j}+ \frac{\partial\bar A_{\mu
j}}{\partial\theta_q}a^*_{\mu j}\right)=
$$
$$
-\frac{1}{2}\eta_{jl}\frac{\partial}{\partial\theta_q}\left(\bar
A^*_{\mu j}\bar A_{\mu l}+\bar A_{\mu j}\bar A^*_{\mu l}\right).
\eqno(12)
$$
Combinations $\frac{\partial\bar A^*_{\mu
j}}{\partial\theta_q}a_{\mu j}+\frac{\partial\bar A_{\mu
j}}{\partial\theta_q}a^*_{\mu j}$ are transverse fluctuations.

Using  $\bar A_{\mu j}$ given by Eq.(11) and taking Fourier
transform of $a_j({\bf r})=d_{\mu}a_{\mu j}({\bf r})$ we arrive at
the following expression for the only finite transverse component
 $a_j({\bf k})$:
$$
l_ja_j({\bf
k})=-\frac{5\sqrt{2}\Delta}{4(5\kappa+\xi_0^2k^2)}[l_j\eta_{jl}({\bf
k})(m_l+in_l)]
   \eqno(13)
$$
 The only non-vanishing average in Eq. (9) originating
 from the transverse fluctuations is:
$$
<a_3(0)a_3^*(0)>=\frac{25}{8}\int\frac{\Delta^2\Phi_0}{(5\kappa+
\xi_0^2k^2)^2}\frac{k^2dk}{\pi^2}=\frac{5\sqrt{5}\Delta^2\Phi_0}{32\pi
\xi_0^3\sqrt{\kappa}}. \eqno(14)
$$
The disorder can be treated as a perturbation when the fluctuation
is small, i.e.
$$
\frac{<a_3(0)a_3^*(0)>}{\Delta^2}\equiv g_{\kappa}\ll 1. \eqno(15)
$$
 With the decreasing $\kappa$ parameter $g_{\kappa}=\frac{5\sqrt{5}\Phi_0}{32\pi
\xi_0^3\sqrt{\kappa}}$ grows as $1/\sqrt{\kappa}$. Perturbation
theory approach breaks down at $g_{\kappa}\sim 1$. At smaller
anisotropy transverse motion of the order parameter can not be
described within the mean field approach.  Situation is analogous
to the critical region in a vicinity of a temperature of a
continuous phase transition,  except that in the case of a weak
quenched disorder only transverse fluctuations are critical.
Longitudinal fluctuations remain small and a short-range order can
be preserved. Intensity of fluctuations is controlled by the
global anisotropy $\kappa$, which in the present case is analogous
to parameter $\tau=(T-T_c)/T_c$ for thermal fluctuations. The
condition $g_{\kappa}\sim 1$ can be used for an order of magnitude
estimation of a borderline anisotropy $\kappa_c$ below which
transverse fluctuations became  critical. Considering aerogel as a
collection of randomly distributed pieces of strand of a length
$\varepsilon$ and of a radius $\rho$ with the average porosity $P$
and using results of the Rainer and Vuorio theory of ``small
objects" in superfluid $^3$He one can obtain the following
estimations \cite{surov}: $\Phi_0\sim \varepsilon\xi_0^2(1-P)$,
$\kappa\sim\gamma(1-P)(\xi_0/\rho)$. Transverse fluctuations are
critical if deformation
$\gamma<\gamma_c\equiv\rho\varepsilon^2(1-P)/\xi_0^3$. For
comparison with the Ref. \cite{vol_conf} let us substitute
$\varepsilon=\xi_a$ as it is assumed there. Here $\xi_a$ is the
average distance between the strands, introduced as
$\pi\rho^2/\xi_a^2=(1-P)$. With this assumption
$\gamma_c\sim(\rho/\xi_0)^3$. When expressed in terms of $\xi_a$
and Larkin-Imry-Ma length $L_{LIM}$ the borderline deformation
$\gamma_c\sim(\xi_a/L_{LIM})^{3/2}$ coincides with the deformation
at which transition from the uniform ABM state to the LIM state is
predicted in Ref.\cite{vol_conf}. It means that the predicted
transition falls into the region where transverse fluctuations are
critical. The mean-field picture used for the prediction of the
transition does not apply in this region and can be used only as a
qualitative guidance. An adequate description of a possible
transition and of the emerging state have to be based on the
formalism used for description of critical phenomena.
Renormalization group analysis of several other systems with a
quenched random anisotropy, in which formation of LIM state would
be expected on a basis of the mean-field argument, proves that a
state with the quasi long-range order (QLRO) forms instead.
\cite{feldm}. In the QLRO state the average order parameter is
zero, but decay of local correlations of the order  parameter with
a distance obeys a power law as it is expected for a decay of
correlations in a critical point.

The order of magnitude estimation of the borderline deformation
for $\rho/\xi_0\sim(1/10)$ yields $\gamma_c\sim 10^{-3}$ as in
Ref.\cite{vol_conf}. Quantitative treatment\cite{surov}  of the
model of strands within the Rainer and Vuorio theory brings this
estimation down to $\gamma_c\sim 10^{-4}\div 10^{-5}$, i.e. a very
high level of isotropy is required for observation of critical
phenomena in the considered system.  But,  as it was pointed out
before \cite{fom1} a deviation of the order parameter of the
A-like phase from the ABM form can start in a region where
transverse fluctuations of the order parameter are still small and
the perturbation theory does apply.

\section {Effect of fluctuations}
For anisotropy $\kappa$ within the interval
$\kappa_c\ll\kappa\ll\tau$ transverse fluctuations are small but
still much greater then the longitudinal: $g_\tau\ll g_\kappa\ll
1$. The estimated critical anisotropy $\kappa_c\sim 10^{-5}\div
10^{-6}$ and $\tau\sim 0.1$, so the interval is wide.
 Within this interval contribution of the
longitudinal fluctuations to Eq. (9) can be neglected. That
simplifies calculation of corrections to the order parameter.
Substitution of expression (14) for fluctuations and Eq. (11) as
the average order parameter  in Eq. (9)  renders an equation for
the gap $\Delta$ of the ABM phase corrected for the transverse
fluctuations:
$$
\tau+\beta_{245}(1+g_{\kappa})\Delta^2=0.      \eqno(16)
$$
It differs from the analogous equation for the unperturbed
ABM-phase by the extra factor $(1+g_{\kappa})$ in front of a sum
of the coefficients $\beta_{245}=\beta_2+\beta_4+\beta_5$.
Parameter $g_{\kappa}$ is positive by its definition. Fluctuations
depress $\Delta^2$ and the condensation energy of the ABM-phase in
comparison with the unperturbed case by a factor
$(1+g_{\kappa})^{-1}$, i.e. the renormalized condensation energy
$f(A_{\mu j}^{ABM})=f_0(A_{\mu j}^{ABM})/(1+g_{\kappa})$.

The amount for which the condensation energy is depressed depends
on a coupling of the average order parameter to the random
anisotropy. There exist a class of orbitally isotropic, or
``robust" order parameters for which the random anisotropy does
not excite transverse fluctuations and there is no ensuing
suppression of their condensation energy ($g_{\kappa}=0$). That
happens when the driving term in the r.h.s. of Eq. (12) vanishes:
$$
\frac{d}{d\theta_q}\left(A_{\mu j}A_{\mu l}^*+A_{\mu l}A_{\mu
j}^*\right)=0.     \eqno(17)
$$
This condition means that the combination in the brackets does not
change at an arbitrary infinitesimal rotation $\theta_q$, i.e.
this combination is proportional to the unit tensor:
$$
A_{\mu j}A_{\mu l}^*+A_{\mu l}A_{\mu j}^*\sim \delta_{jl}.
\eqno(18)
$$
 An immediate example of the robust order parameter is that of BW.
Transverse fluctuations favor robust order parameters over
non-robust. One can conclude that when the global anisotropy is
weak the transverse fluctuations induced by aerogel tend to favor
BW phase over the ABM and to shrink a region of stability of the
ABM phase in comparison with the bulk liquid.

Returning to the A-like phase we have to take into account that it
is an equal spin pairing state.
 Among these states the one satisfying condition (17) up to an arbitrary
rotations in spin and in orbital spaces corresponds to  the A-like
robust order parameter \cite{fom1}:
$$
A^R_{\mu j}=\Delta\frac{1}{\sqrt{3}}[\hat d_{\mu}( m_j+i n_j)+
\hat e_{\mu} l_j],          \eqno(19)
$$
where ${\bf m,n,l}$ are mutually orthogonal orbital unit vectors,
${\bf d,e}$ -- mutually orthogonal unit spin vectors. This order
parameter is not a minimum of the ``bare" free energy $f_0$. The
relative difference of ``bare" energies of the robust and
ABM-states $\varepsilon_0\equiv [f_0(A_{\mu j}^R)-f_0(A_{\mu
j}^{ABM})]/f_0(A_{\mu j}^{ABM})$ can be expressed in terms of the
coefficients $\beta_1,...\beta_5$:
$\varepsilon_0=(\beta_{13}-4\beta_{45})/(9\beta_2+\beta_{13}+5\beta_{45})$.
For the weak coupling values of $\beta$-coefficients this ratio is
1/19, i.e. the density of the `bare" free energy of the robust
state is only slightly higher than that of the ABM-state.  Assume
that the strong coupling corrections to $\beta_1,...\beta_5$ leave
$\varepsilon_0$ small.  The relative difference of renormalized
energies of the two states $\varepsilon\equiv [f(A_{\mu
j}^R)-f(A_{\mu j}^{ABM})]/f(A_{\mu j}^{ABM})$ depends on the
global anisotropy $\kappa$ via parameter $g_{\kappa}$:
$\varepsilon=\varepsilon_0-g_{\kappa}+\varepsilon_0 g_{\kappa}$.
According to Eq. (14) $g_{\kappa}\sim 1/\sqrt{\kappa}$. At
sufficiently small $\kappa$ when
$g_\kappa>\varepsilon_0/(1-\varepsilon_0)$, $\varepsilon<0$ and
the robust state became energetically more favorable than the ABM.
That happens at $g_{\kappa}\approx\varepsilon_0\ll 1$, i.e. the
transverse fluctuations are still small and the perturbation
theory does apply. In terms of a global anisotropy condition
$g_{\kappa}\approx\varepsilon_0$ corresponds to
$\kappa\approx\kappa_c/\varepsilon_0^2\gg\kappa_c$. Comparison of
free energies indicates a possibility of a discontinuous
transition from the ABM into the robust state or in a state with
even lower free energy when the global anisotropy decreases. A
landscape of the free energy of superfluid $^3$He is not yet
established. That impedes a definitive prediction of a character
and position of transition in the robust state. Continuous change
of a form of the order parameter as a function of anisotropy can
not be excluded too. As an illustration of possible changes of a
form of the order parameter consider an interpolation between the
ABM and the robust order parameters:
$$
A^{int}_{\mu
j}=\frac{\Delta}{\sqrt{3+2v^2}}\left[(1-iv)d_{\mu}(m_j+in_j)+e_{\mu}l_j)\right].
   \eqno(20)
$$
At $v\rightarrow \infty$  \quad  $A^{int}_{\mu j}$ goes over into
$A^{ABM}_{\mu j}$ and at $v=0$ -- into $A^R_{\mu j}$. Coefficient
$v$ is a ``fraction" of the ABM-order parameter in $A^{int}_{\mu
j}$. Coupling of the $A^{int}_{\mu j}$ with global  anisotropy is
determined by a combination
$$
A^{int}_{\mu j}(A^{int}_{\mu
l})^*\kappa_{jl}=-\frac{\Delta^2}{3+2v^2}v^2l_jl_l\kappa_{jl}.
\eqno(21)
$$
Coupling with the local anisotropy is obtained by the substitution
of $\eta_{jl}$ instead of $\kappa_{jl}$. For small $v$ both
couplings are weakened by a factor $v^2$. A typical transverse
fluctuation (cf. Eq. (14)) contains $\eta^2$ in the numerator and
$\sqrt{\kappa}$ in the denominator, so that the fluctuation is
proportional to $v^3$. Growth of transverse fluctuations at a
decrease of the global anisotropy $\kappa$ can be compensated by a
choice of sufficiently small $v$ so that the transverse
fluctuations remain small and  region of critical fluctuations is
not entered. The global anisotropy is a convenient parameter for
theoretical analysis. In particular, it makes expressions for
transverse fluctuations finite. The analogy between $\kappa$ and
$\tau=(T-T_c)/T_c$ makes possible to use the theory of critical
phenomena as a guidance. Unfortunately, in practice the anisotropy
(deformation) of aerogel is difficult to control or to vary
continuously. It is particularly difficult in a region of small
deformation $\gamma=\Delta l/l\sim 10^{-2}\div10^{-3}$, which is
of interest. An uncontrolled deformation of such order could be
present in the most of the experiments with $^3$He in aerogel.

\section{Discussion}
Global anisotropy of aerogel lifts continuous degeneracy of
superfluid $^3$He. Direct manifestation of the anisotropy is
orientation of the orbital part of the average order parameter.
Another important effect is a suppression of transverse
fluctuations of the order parameter, which otherwise are critical.
There remain a basic question about the structure of the A-like
phase in the isotropic aerogel. Taking isotropic state as a limit
of vanishing anisotropy  helps to understand its nature.

Different possibilities for a structure of the A-like phase in the
isotropic limit are discussed in the current literature. One of
them  is the LIM state. According to Ref. \cite{vol_conf} it has
to form via a first order phase transition at a certain value of
anisotropy $\kappa$. By the order of magnitude this value
coincides with the borderline anisotropy $\kappa_c$ below which
transverse fluctuations become critical.

Another possibility can be guessed by the analogy with the other
continuously degenerate systems with a quenched random anisotropy.
It is the formation of QLRO state \cite{feldm}. On approach to
 this state when global anisotropy tends to zero the average order
 parameter is presumably fading continuously.
 In both cases only orientation of the order parameter is
 involved. A form of the order parameter does not change.

 In the present paper the third possibility is discussed. It consists
 in the change of a form of the order parameter which decreases its
 coupling with the random
 anisotropy. This adjustment makes possible to maintain a
 long-range order in a limit of vanishing anisotropy. Realization
 of  this possibility in the A-like phase of superfluid $^3$He
 depends on a landscape of the unperturbed free energy $f_0$.
 If the landscape is favorable deviations of the order parameter from
 the ABM form can start at much higher anisotropy then the estimated critical
 value for transition in the  LIM state or in a state with the QLRO.

 No comparison of the expected properties of the proposed state
 with the existing experimental data was made here because of a
 possible ambiguity introduced in the data by an uncontrolled
 deformation of aerogel. One can remark only that neither of the data rules
 out the third possibility.

 Investigation of the A-like phase of $^3$He in aerogel with a possibility
 of tuning deformation of aerogel to a very low level is presently one of the
 most challenging problem in the field.

\begin{acknowledgements}
I am grateful to  V.V. Dmitriev for  useful discussions. This work
is partly supported by RFBR (grant 07-02-00214), Ministry of
Science and Education of the Russian Federation and CRDF (grant
RUP1-2632-MO04).
\end{acknowledgements}


\end{document}